\documentclass[12pt]{article}
\usepackage{amsmath,amssymb,bbold,epsfig}
\usepackage{amsfonts}
\usepackage{bbm}
\usepackage[colorlinks=true,linkcolor=red,citecolor=green,urlcolor=cyan,
bookmarks=true,bookmarksopen=true,pdfpagemode=None,pdfstartview=FitH]{hyperref}
\renewcommand{\vec}[1]{{\mathbf{#1}}}
\newcommand{\appsection}{\addtocounter{section}{1}\setcounter{equation}{0}
                         \renewcommand{\thesection}{\Alph{section}}
}
\renewcommand{\theequation}{\arabic{equation}}
\newcommand{\be}{\begin{equation}}
\newcommand{\ee}{\end{equation}}
\newcommand{\bea}{\begin{eqnarray}}
\newcommand{\eea}{\end{eqnarray}}
\begin{document}

\title{
\vglue -1.5cm
\rightline{\em\normalsize To the memory of G.T.~Zatsepin~~~~~}
\vskip 0.5cm
\Large \bf
Neutrino oscillations: Entanglement, energy-momentum conservation and QFT}
\author{
{E. Kh. Akhmedov$^{a,b}$\thanks{email: \tt 
akhmedov@mpi-hd.mpg.de}~~\,and
\vspace*{0.15cm} ~A. Yu. Smirnov$^{c}$\thanks{email:
\tt smirnov@ictp.it}
} \\
{\normalsize\em $^a$Max--Planck--Institut f\"ur Kernphysik,
Postfach 103980} \\ {\normalsize\em D--69029 Heidelberg, Germany
\vspace*{0.15cm}}
\\
{\normalsize\em $^{b}$National Research Centre Kurchatov
\vspace*{-0.1cm}Institute}\\{\normalsize\em Moscow, Russia 
\vspace*{0.15cm}}
\\
{\normalsize\em $^{c}$The Abdus Salam International Centre for Theoretical    
Physics} \\
{\normalsize\em Strada Costiera 11, I-34014 Trieste, Italy 
\vspace*{0.15cm}}
}
\maketitle 
\thispagestyle{empty} 
\vspace{-0.8cm} 
\begin{abstract} 
\noindent 
We consider several subtle aspects of the theory of neutrino 
oscillations which have been under discussion recently. We show that the 
$S$-matrix formalism of quantum field theory can adequately describe 
neutrino oscillations if correct physics conditions are imposed. This 
includes space-time localization of the neutrino production and 
detection processes.  Space-time diagrams are introduced, which 
characterize this localization and illustrate the coherence issues of 
neutrino oscillations. We discuss two approaches to calculations of the 
transition amplitudes, which allow different physics interpretations: 
(i) using configuration-space wave packets for the involved particles, 
which leads to approximate conservation laws for their mean energies and 
momenta; (ii) calculating first a plane-wave amplitude of the process, 
which exhibits exact energy-momentum conservation, and then convoluting 
it with the momentum-space wave packets of the involved particles. We 
show that these two approaches are equivalent. Kinematic entanglement 
(which is invoked to ensure exact energy-momentum conservation in 
neutrino oscillations) and subsequent disentanglement of the neutrinos 
and recoiling states are in fact irrelevant when the wave packets are 
considered. We demonstrate that the contribution of the recoil particle 
to the oscillation phase is negligible provided that the coherence 
conditions for neutrino production and detection are satisfied. Unlike 
in the previous situation, the phases of both neutrinos from $Z^0$ decay 
are important, leading to a realization of the Einstein-Podolsky-Rosen 
paradox.

\end{abstract}

\vspace{1.cm}
\vspace{.3cm}

\newpage

\section{\label{sec:intro}Introduction}

Although neutrino oscillations appear to be a simple quantum mechanical 
phenomenon, a closer look at them reveals a number of subtle and even 
paradoxical issues. It is probably for this reason that debates on the 
fundamentals of the oscillation theory and on the correctness of different 
theoretical approaches to neutrino oscillations do not cease in the 
literature, with many papers published in the recent years. Some of these 
publications advocate approaches that in our opinion are either incorrect or 
confusing.

Unfortunately, almost every new attempt to revise the basics of the theory 
of neutrino oscillations or to re-interpret the already established results 
is either incorrect or leads to a more complicated and less transparent 
than before formalism, often with many unnecessary details. This hinders  
the understanding of the physics of the oscillation phenomenon and, in 
turn, triggers further incorrect developments. Some confusion 
originates from the fact that there exist several seemingly different 
(though actually equivalent) approaches to neutrino oscillations, 
which allow different physics interpretations. 

In this paper we consider several subtle issues of the theory of 
neutrino oscillations which have been under discussion recently. These  
include inter-related questions of the energy-momentum conservation in 
neutrino oscillations, entanglement of neutrinos and accompanying 
(``recoil'') particles, and the possibility of describing neutrino oscillations 
in the $S$-matrix formalism.  We present our analysis  
in the simplest and most transparent, yet physically adequate way, omitting 
irrelevant details. In the course of our discussion we also comment on some 
incorrect considerations and results which recently appeared in the literature. 
Attempts at implementation in neutrino oscillations of exact energy-momentum 
conservation, kinematic entanglement,  etc., lead to a number of paradoxes, 
for which we present our resolution. In this sense the present  paper can 
be considered 
as a continuation of our previous work ``Paradoxes of neutrino oscillations'' 
\cite{AS1}. 

The paper is organized as follows. In sec.~\ref{sec:pw} we show how neutrino 
oscillations 
can be consistently and adequately described in the $S$-matrix formalism
of quantum field theory (QFT). Sec.~\ref{sec:em} is devoted to the problem of 
energy-momentum conservation in neutrino oscillations in the QFT framework. 
In sec.~\ref{sec:ent} we consider kinematic entanglement and disentanglement 
in neutrino oscillations and show in this connection that the phase of the 
recoil particle  is irrelevant. Here we also discuss relationships between 
neutrino oscillations and the Einstein-Podolsky-Rosen paradox. 
In the Appendices we present our critical comments on some recent papers on 
neutrino oscillations.

\section{\label{sec:pw} $S$-matrix approach of QFT and the physics 
conditions of neutrino oscillations}

Can neutrino oscillations be consistently described in the standard $S$-matrix 
formalism of QFT and, if not, how should this formalism be modified to make 
the description possible? These questions have been addressed in a number 
of publications. Recently, it has been argued \cite{Boy1,Boy2} that 
the $S$-matrix formalism is ill-suited for describing long-baseline 
oscillation experiments because the oscillations are inherently a finite-time 
phenomenon. This is at variance with the results of a number of earlier 
papers, in which it was shown that in the framework of a QFT approach 
with wave packets, finite-distance and finite-time {\em neutrino} evolution 
can be consistently incorporated into the $S$-matrix formalism 
(see~\cite{Beuthe1} for a comprehensive review). Here we present our own  
analysis of this issue and confirm the latter statement. 
Although some points in our discussion may appear rather obvious, the fact 
that their misconception has led to incorrect results shows 
that they deserve a detailed consideration. 

\subsection{\label{sec:physco} Physics conditions of neutrino oscillations }

In principle, QFT and the $S$-matrix formalism  should be able to describe 
any  physical process  in particle physics. 
However, in each particular case the formalism has to 
be adjusted to the specific physical 
situation by making use of, e.g., appropriate initial conditions. Such an 
adjustment is carried out for the processes of scattering of particles,  
particle decays,  {\it etc.}. 
In addition, calculations usually involve certain approximations 
that are process dependent, {\it i.e.}\ are valid for a given situation 
but may not be justified for the other ones. 

\begin{figure}
  \begin{center}
\includegraphics[height=5.2cm,width=12.0cm]{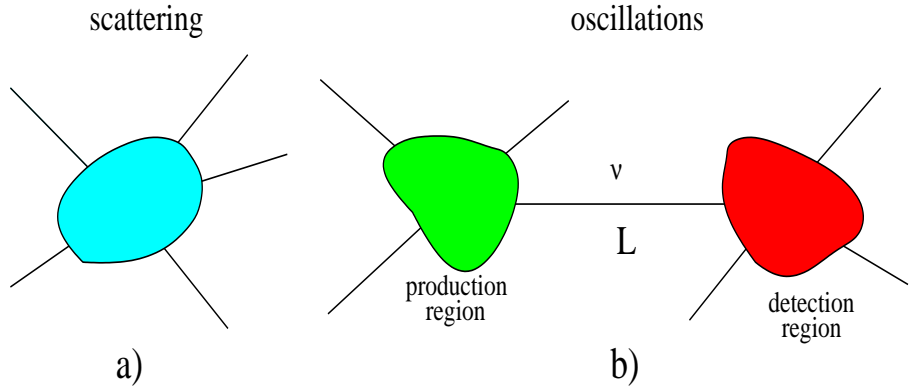}
  \end{center}
\caption{Schematic representation of a scattering process (a),  and 
of a neutrino oscillation setup  
with distinct production and detection regions
           separated by a distance $L$ (b).}
  \label{fig:scheme1}
\end{figure}

Recall that when one employs QFT to describe a scattering process, 
the initial state is considered to represent a system of free non-interacting  
particles. In the process of evolution the particles approach each other, 
enter certain space-time region where they interact, then the products 
of the interaction move apart and after some time are again considered as 
free non-interacting particles. In this setup one deals with a {\it single 
interaction region} (see fig.~\ref{fig:scheme1}a). The following 
approximations are usually made:  

(i) the initial and the final states are considered as asymptotic states 
(defined at $t \rightarrow \pm \infty$) and are described by plane waves; 

(ii) the integration over the 4-coordinate of each interaction point 
is performed over the infinite space-time interval.

\noindent
This results in the proportionality of 
the transition amplitude to $\delta$-functions 
expressing exact energy-momentum conservation in each interaction vertex, 
which substantially simplifies calculations of rates and cross-sections. The 
localization of the source and detector is irrelevant in this setting and is 
not considered. (In fact, the infinite-limits integration and the plane-wave 
description of the asymptotic states means that both the source of the  
particles and the detector are formally taken to be of infinite extension). 

In contrast to this, neutrino oscillations are intrinsically a finite-time 
and finite-distance phenomenon. Neutrinos are produced in a certain confined  
space-time region (the source), then they propagate and are detected in 
another confined space-time region -- the detector. The detector and the 
source have {\it finite} sizes and are separated by a {\it finite} distance 
$L$ (fig.~\ref{fig:scheme1}b). Usually this distance is much larger than the 
sizes of the production and detections regions. The existence in the problem 
of a finite length -- the baseline $L$ -- constitutes a fundamental difference 
from scattering processes. Thus, in the case of neutrino oscillations one deals 
with two distinct interaction regions. (The situation when the production 
and the detection regions overlap or coincide should  be considered 
separately.) The $S$-matrix formalism of QFT should be adjusted 
correspondingly.  

To incorporate the two interaction regions setup in the $S$-matrix formalism 
one should employ appropriate wave functions (wave packets) for 
particles accompanying neutrino production and 
detection (hereafter we will use the term ``external'' for such 
particles).%
\footnote{The name comes from the fact that they correspond to external 
legs of the Feynman diagram describing the overall process of neutrino 
production, propagation and detection, whereas neutrinos are represented 
by the internal line in this diagram (see fig.~\ref{fig:scheme1}b).}
The wave functions of particles which are involved in neutrino 
production ensure the spatial localization of the neutrino 
emission region in the source, whereas the wave functions of particles with 
which neutrinos interact upon propagation should determine the localization 
of the detection region. The approximation of (infinite-extension) plane 
waves for asymptotic states is not valid here -- in that approximation 
the neutrino source and detector cannot be localized, which results in  
oscillations being averaged out. 

Notice that in the case of scattering processes there may exist several 
interaction vertices with 4-coordinates $x_i=(t_i, \vec{x}_i)$. The interaction 
can occur anywhere in the interaction region, and one therefore has to perform 
the integration $\int \prod_i d{\bf x_i} dt_i$ over the whole space-time 
interaction domain for each vertex. This leads to the appearance of 
cross-diagrams. In contrast with this, in the case of neutrino oscillations, 
in the lowest order in weak interactions there are only two interaction 
vertices (one for neutrino production and one for its detection) situated in 
different interaction regions, and the integrations over the corresponding 
4-coordinates $x_1=(t_1, \vec{x}_1)$ and $x_2=(t_2, \vec{x}_2)$ are performed 
over different space-time intervals.  
To a good approximation these regions are spatially separated and time 
ordered. There are no cross diagrams in this case -- the probability that, 
e.g., a particle associated with neutrino production will be born in the 
detector rather than in the source is exponentially suppressed \cite{Boy1}.  
Let us illustrate the above by some examples.

(i) Reactor experiments: $\bar{\nu}_e$ are produced in decays of nuclei $A$ 
and then detected  via the inverse beta decay $\bar{\nu}_e + p \rightarrow 
e + n$. The decaying nuclei are localized in the source (the core of a 
nuclear reactor), and the target protons are localized in the detector. The 
wave function of a parent nucleus $A$ is substantially different from 
zero only in a relatively small region with the size of the order of 
interatomic distances, {\it i.e.}\ much smaller than the size of the 
reactor core. A similar argument applies to the wave functions of target 
protons. It is this localization of the neutrino emitter and receiver that 
determines the positions of the two interaction regions; their sizes are 
given  by the overlap of the wave packets of the particles  
participating in the neutrino production and detection processes 
(see below).

(ii) Accelerator experiments: neutrinos are produced in pion decays, $\pi 
\rightarrow \mu + \nu_\mu$, and the electron neutrinos that appear as a result 
of the oscillations are detected via the $\nu_e + n \rightarrow p + e $ 
reaction, where the neutron is inside a nucleus. Here the pions 
decay in a decay tunnel. They are described by moving wave packets whose 
localization domain may, in fact, be much smaller than the size of the 
tunnel.

\subsection{\label{sec:physco1} Space-time diagrams for oscillations}

\begin{figure}[h]
  \begin{center}
\includegraphics[height=10.2cm,width=14.0cm]{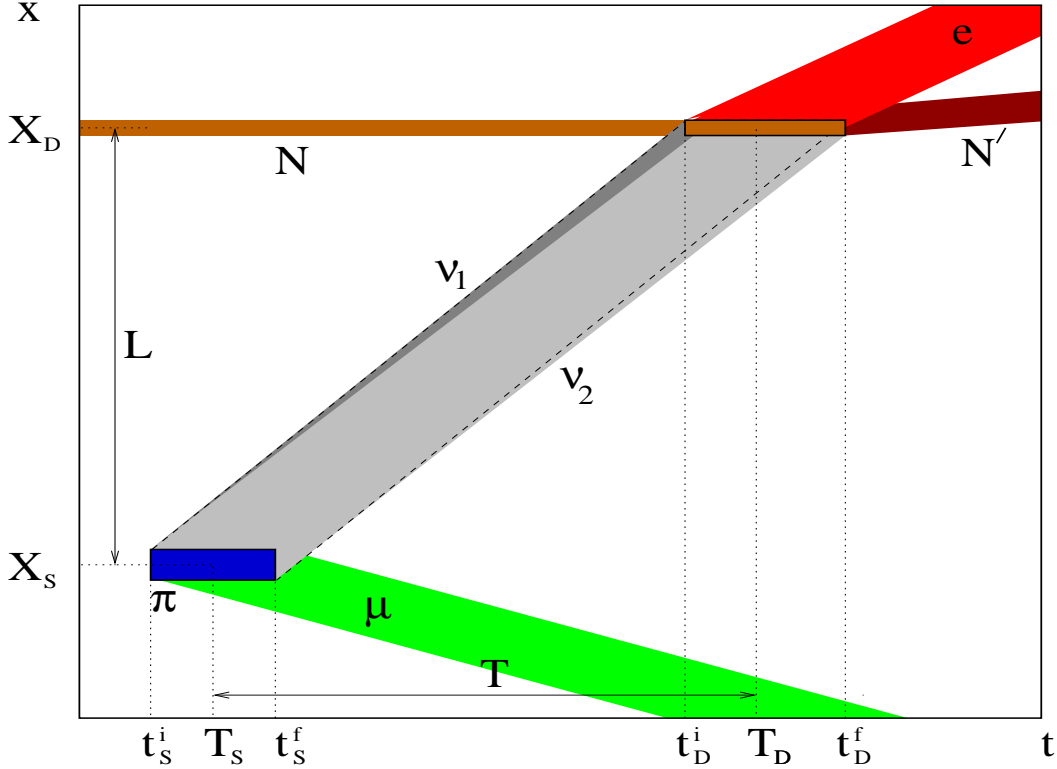}
  \end{center}
\caption{Space-time diagram of neutrino oscillations. Schematic  
representation of neutrino production in pion decay, 
propagation  and observation of the oscillated neutrino  
due to its charged-current interaction with a target nucleon (or nucleus).
See the text for details.}
  \label{fig:scheme2}
\end{figure}

Let us discuss the space-time picture of neutrino production, propagation 
and detection in more detail. For the process described in example (ii),  
the evolution of the system is sketched in fig.~\ref{fig:scheme2}.
The colored  bands correspond 
to  space-time localization  of the participating particles 
as described by wave packets. 
The rectangular regions schematically show the overlap 
domains of the wave packets at neutrino emission and absorption, {\it i.e.}\ 
the neutrino production and detection regions.    
The propagating neutrino state is a superposition of two different mass 
eigenstates, $\nu_1$ and $\nu_2$, moving with slightly different group 
velocities (to make this more clearly seen, we have shown the borders of the 
band corresponding to $\nu_1$ with black dashed lines). If neutrinos propagate 
very long distances, the bands corresponding to different mass eigenstates 
would no longer overlap at the detector, and the oscillations would become 
unobservable.  This corresponds to decoherence due 
to the wave packet separation.
In fig.~\ref{fig:scheme2} the parent pion in the source and the target nucleon 
(or nucleus) in the detector are assumed to be (nearly) at rest. 

Let us elaborate on the space-time characteristics of the interaction regions. 
Since the range of the weak interactions responsible for neutrino production 
and detection is extremely small, the interaction regions are determined by the 
overlap of wave packets of the involved particles. In turn, the shapes and the 
sizes of the wave packets depend on the specific conditions of experiment. 
Notice that the wave packets of the external particles do not have sharp 
borders, and the same applies to the regions of their overlap -- the 
interaction regions.%
\footnote{The bands in fig.~\ref{fig:scheme2} should actually have fuzzy 
peripheries. They are depicted with sharp borders just for simplicity of 
drawing.} 

In process (ii) the space-time domain occupied by the pion 
is determined in the following way. The initial time of this domain (the left 
hand border of the production rectangle in fig.~\ref{fig:scheme2}) is given by 
the time when, e.g., a bunch of protons hits the target, i.e when the pions are 
produced, or when the number of pions in the source is counted. If the pions 
decay at rest in vacuum, the final moment of time of the domain is essentially 
determined by the pion lifetime $\tau_0$. The temporal length of the 
domain can be smaller if, e.g., the decay tunnel is shorter than the pion 
decay length for moving pions. The diagram in fig.~\ref{fig:scheme2} 
corresponds to a situation when the produced muon is not detected and no 
particular space-time conditions are imposed on the detection of the electron 
and recoil nucleus in the detector. In this case the production region is 
determined by the properties of the pion wave packet. The detection region is 
given by the intersection of the bands representing the space-time propagation 
of the wave packets of the neutrino and the neutron.  

The situation can be different if the muon is detected, and a time window 
is imposed on its detection that is shorter than the pion lifetime. This 
narrows down the muon wave packet and consequently the interaction region will 
be determined by the intersection of the pion and muon bands. In this case the 
temporal length of the production region, and consequently, of the detection 
region, will be smaller than the one shown in the fig.~\ref{fig:scheme2}. 
Similarly, the time or space window imposed on the 
electron detection  may narrow down the detection region, and therefore reduce 
the size of the neutrino wave packets, and consequently the production region.

If neutrinos are produced in collision processes, the beginning of the 
interaction region corresponds to the moment of time when the 
wave packets describing the colliding particles begin  to overlap. 

The initial (final) moment of time $t^i_S$  ($t^f_S$) of the production region 
can be defined in such a way that for $t< t^i_S$ ($t > t^f_S$) the overlap of 
the wave packets of particles participating in the neutrino production process 
can be neglected. Recall that the size of the  wave packet of neutrino is 
determined also by the process of neutrino detection. The initial and final 
moments of time of the detection region, $t^i_D$ and $t^f_D$,  can be 
defined similarly.

In fig.~\ref{fig:scheme2} we indicate the 4-coordinates of 
the central points of the neutrino production and detection regions,   
$(T_S, \vec{X}_S)$ and $(T_D, \vec{X}_D)$, respectively.
They can be defined as follows. Let $T_S$ be the time when the overlap of the 
wave packets of particles participating in neutrino production
is maximal; $\vec{X}_S$ is then the position of the central point of the 
overlap region at this time. The time $T_D$ and the coordinate 
of the center of the detection region $\vec{X}_D$ can be defined similarly.  
The baseline $L$ is then given by  the modulus of the vector 
$\vec{L}=\vec{X}_D-\vec{X}_S$,  whereas the mean time elapsed between the 
neutrino production  and detection is $T=T_D-T_S$. These quantities are well 
defined as long as they are large compared to, correspondingly, the spatial 
and temporal extensions of the neutrino production and detection regions.    

Within the $S$-matrix formalism, the calculation of the transition amplitude 
involves the integrations over 4-coordinates of the neutrino production and 
detection points $x_1$ and $x_2$ which  are performed over the infinite 
space-time interval.
The transition amplitude corresponding to the production and detection 
of the $i$th neutrino mass eigenstate can be written as 
\be
{\cal A}_i =
\int d^4 x_1  \int d^4 x_2
\prod_{j} \Psi_j (x_1, {\bf X}_S, T_S)
\prod_{l} \Psi_l (x_2, {\bf X}_D, T_D) M_i(x_1, x_2)\,.
\label{eq:A1}
\ee
Here $j$ ($l$) runs over the external particles interacting in the production 
point $x_1$ (detection point $x_2$), $\Psi_j (x_1, {\bf X}_S, T_S)$ and 
$\Psi_l (x_2, {\bf X}_D, T_D)$ are the wave functions (wave packets) of the 
external particles in the case of incoming particles or complex conjugates 
of the wave functions in the case of the outgoing particles and, as we 
discussed, ($T_S$, ${\bf X}_S$), ($T_D$, ${\bf X}_D$) 
determine the space-time location of the interaction regions. 
The quantity $M_i(x_1,x_2)$ is the remainder of the transition matrix 
element. The amplitude of the overall process of $\nu_\alpha$ production, 
propagation (including the oscillations) and $\nu_\beta$ detection 
is then given by the sum 
\be
{\cal A}^{tot}_{\alpha\beta} = \sum_i U_{\alpha i}^* U_{\beta i} {\cal A}_i\,,
\label{eq:ampl}
\ee
where $U$ is the leptonic mixing matrix. The relative phases between the 
amplitudes ${\cal A}_i$ with different $i$ determine the oscillation phases. 

The dominant contributions to the integrals in (\ref{eq:A1}) come from finite 
space-time intervals which correspond to the production and detection domains 
discussed above. In particular, the infinite-time integrations can be 
replaced by those over finite time intervals: $t_i^S\le t_1\le t_f^S$ and 
$t_i^D\le t_2\le t_f^D$, or by any time intervals encompassing them  
-- the result will be practically independent of the integration 
limits provided that they include the interaction regions.

Alternatively, one can still use the plane wave approximation for the wave 
functions of the external particles.  
But in this case, in order to ensure proper localization of the neutrino 
production and detection processes, one should impose ``by hand'' the  finite 
integration limits for the 4-coordinates $x_1$ and $x_2$,  
corresponding to the finite space-time domains of the interactions. 
In eq.~(\ref{eq:A1}) finite {\em effective} integration 
intervals were selected automatically. The plane-wave description plus 
imposition of finite integration limits can be considered as an approximation 
to the localized wave functions description of neutrino oscillations with  
infinite-limits integrations over $x_1$ and $x_2$. Using such 
``finite-extension plane waves'' is sometimes invoked to argue that the plane 
wave approach still gives a viable description of neutrino oscillations. It 
should be remembered, however, that a finite ``piece'' of a plane wave is no 
longer a plane wave -- it is a wave packet.%
\footnote{Indeed, a 1-dimensional ``piece'' of the spatial size $2 a$ of a 
plane wave with momentum $p_0$ is actually a wave packet with the momentum 
distribution function $f(p,p_0)=const\times \sin[(p-p_0) a]/(p-p_0)$ rather 
than $f(p,p_0)=\delta(p-p_0)$, which would correspond 
to a true (infinite-length) plane wave.}
Any consistent description of neutrino oscillations must therefore employ wave 
packets for the external particles (or at least for some of them). The plane 
wave approach cannot describe the localization of the neutrino source and 
detector, without which neutrino oscillations would be unobservable.

Since the neutrino source and detector are of finite size, in realistic 
situations one should, in fact, both use wave packets and perform 
finite-interval integrations. However, if the limits imposed by 
macroscopic sizes of the source and detector are large compared to the 
(usually microscopic) space-time regions of the {\it individual} neutrino 
production and detection processes, the results obtained with finite-interval 
integrations practically coincide with those corresponding to 
infinite-limits integration.  

\subsection{\label{sec:fintime}Finite-limits time integration and the 
observation time}

In refs.~\cite{Boy1,Boy2} a finite evolution time approach to neutrino 
oscillations has been developed, and expressions for the oscillation 
probabilities that differ from the standard one were obtained. The results 
of~\cite{Boy1,Boy2} follow from the 
authors' use of incorrect physics conditions that do not correspond to a 
realistic setup of neutrino oscillations. In \cite{Boy1,Boy2} all the 
particles involved in neutrino production and detection as well as the 
neutrinos themselves were described by plane waves, and the integration 
over the detection time $t_2$ was performed over the interval which extends 
from the time of observation of the charged lepton accompanying the neutrino 
emission in the source to the time of observation of the charged lepton born 
in the neutrino absorption process in the detector. These instants of time 
approximately coincide with the times when the neutrino production and 
detection processes end, {\it i.e.}\ with $t^f_S$ and $t^f_D$ in our notation.  
In other words, in the analysis of \cite{Boy1,Boy2} the detection region 
was immediately attached to the production region.  

While integrating over a very long detection time interval would be perfectly 
admissible in the wave packet description of neutrino oscillations (the wave 
packets overlap would automatically select the correct interaction region), 
this leads to grave consequences in the plane-wave framework adopted 
in~\cite{Boy1,Boy2}. As we pointed out in secs.~\ref{sec:physco} and 
\ref{sec:physco1}, the plane wave approximation 
augmented by judiciously selected finite integration intervals could to some 
extent imitate the wave packet approach. This, in particular, means that the 
integrations should be performed over the production and detection regions 
(the rectangles in fig.~\ref{fig:scheme2}). The approach of~\cite{Boy1,Boy2}, 
in which the integrations were extended from $t^f_S$ to $t^f_D$, actually 
amounts to assuming the neutrino detector to be as long as the experimental 
baseline itself (see Appendix~A for a more detailed discussion). 

In refs.~\cite{Boy1,Boy2} the authors confused 
the total {\it time of experimental observation} $t_{exp}$ (which they call 
``the total reaction time'') with the neutrino evolution time $T$, which is 
the mean interval of time between the neutrino production and detection. 
Their expressions for the oscillation 
probabilities exhibit both proportionality to $T$  and an oscillatory behavior 
in $T$. The correct calculations give the proportionality of the overall 
probability of the process as well as of the event numbers to $t_{exp}$, and 
an oscillatory behavior in $T$~\cite{AK1}. 

In this connection, let us clarify the meaning of the time of the experimental 
observation $t_{exp}$ and its connection to the time intervals introduced in 
secs.~\ref{sec:physco} and \ref{sec:physco1}. In practice, $t_{exp}$ is the 
time interval during which measurements are performed,  
e.g. the time interval between the moments when the detector is turn on
and off, or the time during which the accelerator provides a beam of 
particles that produce neutrinos. Thus, $t_{exp}$ is essentially the time 
of continuous accumulation of statistics (at neutrino detection) or 
luminosity (if we speak about the neutrino production). 
This time can range from a very small fraction of second, if a 
special time window for observation is imposed (as, for instance, in K2K 
and MINOS accelerator experiments), to years, when, e.g., an underground 
detector accumulates neutrino events.

The dependence of the number of events on $t_{exp}$ and possible relations 
between $t_{exp}$ and the detection and production time intervals are 
determined by the nature of the external states involved in the process. 

1.  In the case of stationary external states which, in turn, produce
stationary neutrino states (e.g., neutrino production in decays of very long 
lived isotopes) the time $t_{exp}$  is directly related to the detection 
time: $t_{exp}= t_D^f - t_D^i$. It is assumed here that the detection 
process is also stationary. The time $t_{exp}$ determines the size of the wave 
packet of, e.g., the electron in the final state of the example shown in 
fig.~\ref{fig:scheme2} (once the wave packet of the neutron is given), and 
consequently, the temporal extension of the detection region. 
Therefore, $t_{exp}$ determines also the limits of integration over times 
$t_1$ and $t_2$ in (\ref{eq:A1}). The length of the integration intervals 
is identified with the time of experimental observation.
If the observation time is much larger than the neutrino propagation time,  
$t_{exp} \gg T$, the quantity $t_{exp}$ determines approximately the total 
time of the production-propagation-detection process. The number of 
detected events is then simply proportional to $t_{exp}$. Mathematically, this 
proportionality is a consequence of the proportionality of the transition 
amplitude to the energy-conserving $\delta$-function $\delta(E_f-E_i)$, which 
appears due to the {\it stationary situation} \cite{Grimus:1996av}. 
Indeed, for large $t_{exp}$ the probability is proportional to  
\be 
[\delta(E_f-E_i)]^2 \propto t_{exp}\,\delta(E_f-E_i). 
\ee

2. The situation is different when the external particles are described 
by relatively short moving wave packets, so that the lengths of the interaction 
intervals $(t^i_S, t^f_S)$ and $(t^i_D, t^f_D)$, which are determined by 
the overlap times of the corresponding external wave packets, are much 
smaller than $t_{exp}$ and $T$. The amplitude of the 
process is not proportional to the energy-conserving $\delta$-function 
in this case~\cite{Beuthe1}. Therefore, the probability of the {\it individual} 
neutrino production-propagation-detection process is 
not proportional to the time of observation $t_{exp}$ even in the $S$-matrix
formalism, contrary to the claim in \cite{Boy1,Boy2}. This is because such a
proportionality arises only in a stationary situation, whereas the process
with production and detection of a single neutrino is highly non-stationary.
The proportionality to $t_{obs}$ can appear, however, 
if one deals with stationary beams of incoming particles or ensembles
of emitters rather than with individual processes
(see \cite{AK1}, sec. 5.2). In this case the intervals of integration over 
$t_1$ and $t_2$ in (\ref{eq:A1}) are not related to $t_{exp}$; in 
particular, they can be much smaller than $t_{exp}$.

In any case, the number of events should not oscillate with $t_{exp}$. 
It should,  however, oscillate with $T$ -- the time elapsed between 
the production and detection of neutrinos -- if the $T$- and $L$-dependent 
probability of the process $P_{\alpha\beta}(T, L) = 
|{\cal A}^{tot}_{\alpha\beta}(T, L)|^2$ is considered. 
Note that since the neutrino production and detection times are usually not 
measured (or at least not measured accurately enough), 
the probability $P_{\alpha\beta}(T, L)$ 
is often integrated over $T$. However, this integration is not equivalent to 
the one in \cite{Boy1,Boy2} and it does not reproduce the results of those 
papers. Indeed, $P_{\alpha\beta}(T, L)$ is 
significantly different from zero only if $T\simeq L/v$, 
where $v$ is the average group velocity
of the neutrino state, with possible deviations from the exact equality given
by the length of the neutrino wave packet $\sigma_x$.%
\footnote{For example, in the case of Gaussian wave packets one finds
$P_{\alpha \beta}(T, L) \propto \exp\Big[-\frac{(L-v T)^2}{2\sigma_x^2}\Big]$
\cite{Giun1,Beuthe1}.} In the limit $\sigma_x \rightarrow 0$, the probability
becomes proportional to $\delta (L - v T)$, so that the integration simply 
identifies $v T$ with $L$. For finite $\sigma_x$, the integration over $T$  
effectively amounts to averaging the probability over the finite sizes of the 
production and detection regions. 
In any case, this integration yields the oscillation probability 
$P_{\alpha\beta}(L)$ which exhibits an oscillatory behavior only in 
$L$.

\section{\label{sec:em}
Energy-momentum conservation and neutrino \\oscillations}

\subsection{\label{sec:em1}Energy-momentum conservation and localization}

As has been pointed out by many authors, the assumption of exact energy and 
momentum conservation, if immediately  
applied to particles participating in the 
neutrino production and detection processes, would make neutrino oscillations 
impossible. The reason for this is twofold. First, exact energy-momentum 
conservation would allow one to determine the energy $E_\nu$ and momentum 
$\vec{p}_\nu$ of neutrino from those of the external particles, provided that 
energy and momenta of these particles are known precisely. 
Since neutrinos propagate macroscopic distances and are therefore on the mass 
shell, one can then determine the neutrino mass from the relation $E^2_\nu = 
\vec{p}^2_\nu + m^2$. This would mean that the neutrino state is a mass 
eigenstate rather than a coherent superposition of different mass eigenstates. 
Note that mass eigenstates do not oscillate in vacuum. 
Second, exact energy-momentum conservation implies that all the involved 
particles have sharp energies and momenta, {\it i.e.}\ are described by plane 
waves. This would make localization of the neutrino source and detector 
impossible, leading to a washout of neutrino oscillations. The former 
difficulty could to some extent be alleviated through kinematic 
entanglement~\cite{Goldman:1996yq,Dolgov,Nauen,Cohen1} (see the next section), 
but the latter one remains.

{\it As has been stressed in \cite{AS1}, 
this does not, of course, mean that energy-momentum conservation, which 
is a fundamental law of nature, is violated: it is satisfied exactly when one 
applies it to all particles in the system, including those whose interactions 
with the particles directly involved in the process in question localize the 
latter in given space-time regions. In other words, energy and momentum 
uncertainties are not in contradiction with energy-momentum conservation, 
which is exact for closed systems. }

In most situations (e.g.,\ in scattering experiments), 
the contributions of the localizing particles to the overall 
energy-momentum balance of the system are not taken into account. Since 
the resulting inaccuracy of the energy and momentum conservation, as well as 
the intrinsic quantum-mechanical uncertainties of the energies and momenta of 
the involved particles, are completely negligible compared to their 
energies and momenta themselves, they can be safely ignored in most 
processes. This is, however, not justified when neutrino oscillations are 
considered, since in the absence of neutrino energy and momentum 
uncertainties the oscillations just would not occur. Therefore attempts 
to base the analyzes of neutrino oscillations on exact energy-momentum 
conservation are inconsistent.  

As we shall show now, exact energy-momentum conservation can play a role at an  
intermediate stage of QFT calculations of the transition amplitudes, when one 
deals with plane waves. But at this stage the oscillations cannot be obtained.

\subsection{\label{sec:QFTem}
QFT and energy-momentum conservation}

There are two approaches to the calculation of the amplitude of 
the neutrino production-propagation-detection process. In the first approach 
the external particles are described from the outset by localized wave packets 
in the configuration space, whereas in the second one the transition amplitude 
is first calculated in the plane-wave approach and then is convoluted with the 
momentum distribution functions of the external particles which characterize 
the momentum spreads of their wave packets and take into account their 
localization. As we shall show, the two approaches are actually equivalent; 
however, they allow different physics interpretations, in particular, 
regarding the role of energy-momentum conservation in neutrino oscillations. 
Let us consider these approaches in turn.  

1. The external particles are described by localized configuration-space wave 
functions (wave packets).  The transition amplitude corresponding to the 
production and detection of the $i$th neutrino mass eigenstate $\nu_i$   
is given in (\ref{eq:A1}). 
The quantity $M_i(x_1, x_2)$ in this equation is proportional to the 
propagator of $\nu_i$. Due to invariance with respect to space-time 
translations, $M_i(x_1, x_2) = M_i(x_2-x_1)$, and therefore it 
can be represented as a Fourier integral
\be
M_i(x_2-x_1)=\int \frac{d^4q}{(2\pi)^4}\, M_i(q) e^{-i q (x_1-x_2)}. 
\label{eq:M1}
\ee
The wave functions $\Psi_j (x, {\bf X}_S, T_S)$ 
can also be represented as Fourier transforms of the corresponding 
momentum-space wave functions $f_j(\vec{p},\bar{\vec{p}}_j)$:
\be
\Psi_j (x, {\bf X}_S, T_S)= \int\frac{d {\bf p}}{(2\pi)^3}
f_j(\vec{p},\bar{\vec{p}}_j) 
e^{-i \varepsilon_j E_j(\vec{p})(t-T_S)+i\varepsilon_j 
\vec{p}(\vec{x}-\vec{X}_S)}\,.
\label{eq:Four1}
\ee
Here $\varepsilon_j=+1$ stands for the initial-state particles and $-1$ for the 
final-state ones, and $f_j(\vec{p},\bar{\vec{p}}_j)$ is the momentum 
distribution function of the $j$th external particle participating in neutrino 
production. It is characterized by a peak which is centered at $\vec{p}=
\bar{\vec{p}}_j$ and has the width $\sigma_{pj}\sim \sigma_{xj}^{-1}$, where 
$\sigma_{xj}$ is the spatial length of the wave packet $\Psi_j$. The latter 
follows from Heisenberg's uncertainty relation, and at the mathematical level 
is a reflection of the well-known property of the Fourier transformation. 

The choice of the $t$- and $\vec{x}$-independent phase factor 
($e^{i\varepsilon_j [E_j(\vec{p})T_S-i\vec{p}\vec{X}_S]}$) in the 
integrand of eq.~(\ref{eq:Four1}) corresponds to the condition that the peak 
of the wave packet in the configuration space 
is at the point $\vec{x}=\vec{X}_S$ at the time $t=T_S$ 
(indeed, away from the point $(t,\vec{x})=(T_S, \vec{X}_S$) the integrand 
becomes fast oscillating, leading to a strong suppression of the integral). 
This phase factor can be absorbed into a redefinition of the momentum 
distribution function, {\it i.e.}\ $f_j(\vec{p},\bar{\vec{p}}_j)\to
\tilde{f}_j(\vec{p},\bar{\vec{p}}_j; T_S,\vec{X}_S)$: 
\be
\tilde{f}_j(\vec{p},\bar{\vec{p}}_j; T_S,\vec{X}_S)
=f_j(\vec{p},\bar{\vec{p}}_j)
e^{i\varepsilon_j[E_j(\vec{p})T_S-i\vec{p}\vec{X}_S]}\,.
\label{eq:tildaf1}
\ee
When expressed through 
$\tilde{f}_j(\vec{p},\bar{\vec{p}}_j; T_S,\vec{X}_S)$, the 
Fourier integral in~(\ref{eq:Four1}) takes a more familiar form, with 
the phase factor 
$e^{-i\varepsilon_j[E_j(\vec{p})t - \vec{p}\vec{x}]}\equiv e^{-i\varepsilon_j 
p_j x}$ in the integrand. In addition to the information on the mean momentum 
and the width and shape of the peak, the momentum distribution function 
$\tilde{f}_j(\vec{p},\bar{\vec{p}}_j; T_S,\vec{X}_S)$ encodes the information 
on the location of the center of the wave packet in the configuration space at 
a fixed time $T_S$. 

Similar Fourier expansions and similar considerations apply also to the 
wave functions of the external particles participating in neutrino 
detection. Substituting these expansions as well as eqs.~(\ref{eq:Four1}) 
and~(\ref{eq:tildaf1}) into expression~(\ref{eq:A1}) for the transition 
amplitude, we find 
\bea
{\cal A}_i = 
\int d^4 x_1 d^4 x_2
\prod_{j}\int \frac{d\vec {p}_j}{(2\pi)^3} 
\tilde{f}_j(\vec{p}_j,\bar{\vec{p}}_j; T_S,\vec{X}_S)
\prod_{l} 
\int \frac{d \vec{p}_l}{(2\pi)^3} 
\tilde{f}_l(\vec{p}_l,\bar{\vec{p}}_l; T_D,\vec{X}_D) \nonumber \\
\times \,M_i(x_2-x_1)\,
e^{-i\varepsilon_j p_j x_1 -i\varepsilon_l p_l x_2} \,.
\label{eq:A1a}
\eea
{}From this equation one can see that the mean values of the energies 
and momenta of the involved particles satisfy only 
{\it approximate conservation 
laws}. Indeed, if the momentum distribution functions $\tilde{f}_j(\vec{p}_j,
\bar{\vec{p}}_j; T_S,\vec{X}_S)$ and $\tilde{f}_l(\vec{p}_l,\bar{\vec{p}}_l; 
T_D,\vec{X}_D)$ were proportional to, respectively, $\delta^{(3)}(\vec{p}_j
-\bar{\vec{p}}_j)$ and $\delta^{(3)}(\vec{p}_l-\bar{\vec{p}}_l)$, 
the integrations over $d \vec{p}_j$  and $d\vec{p}_l$ would 
be removed by these $\delta$-functions, and the 
remaining integral over $d^4 x_1 d^4 x_2$ of the expression in the second line 
of eq.~(\ref{eq:A1a}) would result in the standard energy-momentum conserving 
$\delta^{(4)}$-function. In reality, the momentum distribution functions are 
not of $\delta$-type, but they are peaked at the corresponding mean 
momenta, with the peak widths $\sigma_{p}\ll |\bar{\vec{p}}|$. 
Therefore the mean energies and momenta of the particles involved in the 
neutrino production and detection processes satisfy only approximate 
conservation laws, with deviations from exact conservation determined 
by the values of the widths $\sigma_{p}$ of the momentum distribution 
functions. 
 
2. Alternatively, one can first calculate the transition amplitude in 
the plane-wave approach ({\it i.e.},\ considering the external particles to 
be described by plane waves), and then convolute the obtained result with 
the actual momentum distribution functions of the external particles: 
\be
{\cal A}_i = 
\prod_{j}\int \frac{d \vec{p}_j}{(2\pi)^3} 
\tilde{f}_j(\vec{p}_j,\bar{\vec{p}}_j; T_S,\vec{X}_S)
\prod_{l} \int \frac{d \vec{p}_l}{(2\pi)^3} 
\tilde{f}_l(\vec{p}_l,\bar{\vec{p}}_l; T_D,\vec{X}_D)\,
{\cal A}_i^{pw}(\{p_j\},\{p_l\})\,. 
\label{eq:A2} 
\ee
The plane-wave amplitude of the process,  
${\cal A}_i^{pw}(\{p_j\},\{p_l\})$,  is given by
\be
{\cal A}_i^{pw}(\{p_j\},\{p_l\})=\int d^4 x_1 d^4 x_2\,M_i(x_2-x_1)\,
e^{-i(\sum_j \varepsilon_j p_j)  x_1-i(\sum_l\varepsilon_l p_l) x_2} \,.
\label{eq:pw}
\ee 
This amplitude is proportional to a 
$\delta^{(4)}$-function expressing the 4-momentum conservation in the 
process with external particles represented by plane waves. Indeed, 
substituting into~(\ref{eq:pw}) the Fourier expansion (\ref{eq:M1}) for 
$M_i(x_2-x_1)$ and performing the integration over $q$, we find  
\be
{\cal A}_i^{pw}(\{p_j\},\{p_l\})\propto\,
\delta^{(4)}\left(\sum_j \varepsilon_j p_j+\sum_l\varepsilon_l 
p_l\right)\,.
\label{eq:delta}
\ee
The amplitudes ${\cal A}_i^{pw}$, if substituted into eq.~(\ref{eq:ampl}), 
would not lead to neutrino oscillations. This is related to the fact 
that plane waves are completely delocalized in space and time. 
The convolution in eq. (\ref{eq:A2}) takes into account that the external 
particles are actually localized, and it this localization  
that actually makes the oscillation possible. 

It is easy to see that the described two approaches to the calculation of 
the transition amplitude are equivalent.   
Indeed, substituting (\ref{eq:pw}) into (\ref{eq:A2}) yields
\bea
{\cal A}_i = 
\prod_{j}\int \frac{d \vec{p}_j}{(2\pi)^3} 
\tilde{f}_j(\vec{p}_j,\bar{\vec{p}}_j; T_S,\vec{X}_S)
\prod_{l} \int \frac{d \vec{p}_l}{(2\pi)^3} 
\tilde{f}_l(\vec{p}_l,\bar{\vec{p}}_l; T_D,\vec{X}_D)\,
\nonumber \\
\times \int d^4 x_1 d^4 x_2\,M_i(x_2-x_1)\,
e^{-i \varepsilon_j p_j  x_1-i\varepsilon_l p_l x_2} \,.
\label{eq:A2a} 
\eea
The expressions for the amplitudes in eqs.~(\ref{eq:A1a}) and~(\ref{eq:A2a}) 
coincide. Indeed, one of them is immediately obtained from the other by 
interchanging the order of the integrations over the momenta and 
4-coordinates, $d^4 x_1 d^4 x_2$ and $\prod_{j,l} d \vec{p}_j d \vec{p}_l$. 

The two expressions for the transition amplitude allow for different physical 
interpretations. The amplitude in eq.~(\ref{eq:A1a}) is calculated with 
the coordinate-dependent localized wave functions of the external particles. 
{}From its construction it follows that the energies and momenta of the 
involved particles are not sharp, and their mean values satisfy only 
approximate conservation laws, as was discussed above. This is essential for 
neutrino oscillations to occur: indeed, as we have already discussed, too 
accurate measurements of neutrino energy and momentum would destroy neutrino 
oscillations.

On the other hand, the second approach emphasizes the relation of the 
calculation with the fact that energy and momentum conservations are actually  
exact laws of nature. It can be readily seen from (\ref{eq:A2}) and 
(\ref{eq:delta}) that the {\em individual} momentum components of the wave 
packets of the external particles (corresponding to the plane-wave amplitudes) 
satisfy exact energy and momentum conservation laws. The fact that the 
external particles are localized and conservation of their mean energies and 
momenta has only approximate nature is properly taken into account when one 
convolutes the plane-wave transition amplitude given by the second line of 
eq.~(\ref{eq:A2a}) with the momentum distribution functions $\tilde{f}_j$ and 
$\tilde{f}_l$, as is expressed by the outer integrals in~(\ref{eq:A2a}). This 
is in line with the well known fact that, while plane waves are fully 
delocalized, integrating them over small intervals $\sim \sigma_p$ of momenta 
leads to constructive interference in certain space-time intervals of 
width $\sigma_x\sim 1/\sigma_p$ and destructive interference everywhere else.

\section{\label{sec:ent} Entanglement and the recoil phase}

In refs. \cite{Goldman:1996yq,Nauen,Cohen1,Boy1,Hamish} it has been argued 
that exact energy-momentum  conservation can still be applied to description 
of neutrino oscillations. This exact conservation leads to a kinematic 
entanglement of neutrinos with accompanying charged leptons.  
According to ~\cite{Nauen,Cohen1,Boy1}, the subsequent disentanglement 
is assumed to be due to the detection of the charged leptons or 
due to their interaction with the environment. This localizes the charged 
leptons and creates energy and momentum uncertainties for 
the neutrino state, which are necessary for the oscillations to 
occur. The disentanglement processes then has to be explicitly included in 
the description of neutrino oscillations. In what follows we will explore the 
relevance and physical meaning of this ``entanglement/disentanglement'' 
approach.  
 
\subsection{\label{sec:kinent}Kinematic entanglement and coherence}

For definiteness we will speak about the $\pi \to \mu \nu$ decay, 
keeping in mind the general case of two-body decay $P \rightarrow R + 
\nu$, where $P$ and  $R$ are the parent and recoil particles, respectively. 

Suppose first that the 4-momentum of the pion $p_\pi$ is well defined.   
Then, exact energy-momentum conservation would mean that, for each 
emitted neutrino mass eigenstate $\nu_i$ with a   
certain 4-momentum $p_{\nu i}$, the 4-momentum $p_{\mu i}$ of the 
accompanying muon must satisfy  
\be
p_{\nu i} + p_{\mu i} = p_\pi, ~~~ i = 1, 2 
\label{eq:exact}
\ee
(for simplicity we consider 2-flavor mixing here).  
In this way, the state produced in the pion decay can be a coherent 
superposition of different neutrino mass eigenstates accompanied by the muon 
states, 4-momenta of which are correlated with those of neutrinos 
(hence the name entangled state):
\be
|\mu\,\nu \rangle =\sum_i U_{\mu i}^* |\mu(p_{\mu i})\rangle 
|\nu_i(p_{\nu i})\rangle. 
\label{eq:entang}
\ee
If the 4-momentum of the muon is measured very accurately  
and the result gives, e.g., $p_{\mu 1}$, then the neutrino detector should 
observe only $\nu_1$ with the 4-momentum $p_{\nu 1}$. This is a realization 
of the Einstein-Podolsky-Rosen (EPR) correlation \cite{EPR}. In this case, 
however, no interference and therefore no oscillations would occur: 
Measuring the muon momentum with precision higher than 
$|{\bf p}_{\mu 1} - {\bf p}_{\mu 2}|$ would destroy the coherence of 
the components of the entangled state corresponding to different neutrino 
mass eigenstates.  

As has been mentioned above, disentanglement amounts to a measurement 
of the muon momentum with a sufficiently large intrinsic uncertainty (due 
to a sufficiently good localization of the measurement process). Therefore, 
it implies a violation of the strict correlation (\ref{eq:exact}) between the 
muon and neutrino 4-momenta, that is, it leads to a separation of the muon 
and neutrino parts of the state (\ref{eq:entang}). On the other hand, 
coherence loss is a splitting of the components of the state (\ref{eq:entang}) 
which correspond to different neutrino mass eigenstates. We will show here 
that there actually exists an intimate relationship between  the coherence 
and entanglement/disentanglement in the context of neutrino oscillations. 

According to (\ref{eq:ampl}), the transition amplitude describing the  
oscillation process is the sum of the amplitudes which correspond to 
the emission and detection of $\nu_1$ and $\nu_2$:
\be
{\cal A}^{tot}_{\alpha \beta} = U_{\alpha 1}^* U_{\beta 1}^{}
{\cal A}_1 (p_{\mu 1}, p_{\nu 1}) + 
U_{\alpha 2}^* U_{\beta 2}^{} {\cal A}_2 (p_{\mu 2}, p_{\nu 2})\,. 
\ee
Consequently, the probability of the process, 
$|{\cal A}^{tot}_{\alpha \beta}|^2$, contains the interference term 
which is the real part of the expression 
\bea
U_{\alpha 1}^{} U_{\beta 1}^* U_{\alpha 2}^* U_{\beta 2}^{} 
\int_{D_\mu} d{\bf x}_\mu 
{\cal A}_1 (p_{\mu 1}, p_{\nu 1})^*  
{\cal A}_2 (p_{\mu 2}, p_{\nu 2})
\nonumber \\
\propto   
U_{\alpha 1}^{} U_{\beta 1}^* U_{\alpha 2}^* U_{\beta 2}^{}    
\int_{D_\mu} d{\bf x}_\mu 
e^{i (E_{\mu 1}  - E_{\mu 2}) t}
e^{i ({\bf p}_{\mu 2}  - {\bf p}_{\mu 1}){\bf x}_{\mu}}\,,~
\label{eq:interf}
\eea
which is responsible for the oscillations.  Here the integration is over the 
muon detection region ${D_\mu}$. If the size of the detector is smaller than 
$|{\bf p}_{1 \mu} - {\bf p}_{2 \mu}|^{-1}$, the integral is significantly 
different from zero and neutrino oscillations are observable; in the opposite 
case the integral is strongly suppressed due to the fast oscillations of the 
integrand, and the oscillations are averaged out. 

In reality, the 4-momentum of the parent pion is never fixed precisely. 
The very fact that the pion decays implies that it is localized in space and 
time and hence its energy and momentum have uncertainties.  
Therefore, the pion must be described by a wave packet characterized by a 
momentum distribution function of a width $\sigma_{\pi p}$. 
This means that there is no strict correlation between the 4-momenta of the 
neutrino and muon produced in the pion decay. For a given value $p_{\nu i}$ of 
the 4-momentum of the $i$th neutrino mass eigenstate, the muon 4-momentum 
$p_{\mu i}$ is no longer uniquely determined by eq.~(\ref{eq:exact}); 
instead, it can take any value within a range of width of the order 
$\sigma_{\pi p}$. In other words, now instead of eq.~(\ref{eq:exact}) we 
have 
\be
p_{\nu i} + p_{\mu i} = p_{\pi i}, ~~~ i = 1, 2\,,
\label{eq:twoeq}
\ee
where $p_{\pi i}$ is no longer uniquely fixed. 
Consider, for example, the case when the 4-momenta of the components 
of the muons wave function  
accompanying the different neutrino mass eigenstates coincide, {\it i.e.}\ 
$p_{\mu 1} = p_{\mu 2}\equiv p_\mu$. This situation is realized when the muon 
is not detected, which corresponds to the limit $D_\mu \to \infty$, so that 
the expression in eq.~(\ref{eq:interf}) is proportional to $\delta^{(3)}
(\vec{p}_{\mu 1}-\vec{p}_{\mu 2})$. One can then write 
\be 
{\bf p}_\mu + {\bf p}_{\nu 1} = {\bf p}_{\pi 1}\,,\qquad\quad 
{\bf p}_\mu + {\bf p}_{\nu 2} = {\bf p}_{\pi 2}\,. 
\label{eq:mom} 
\ee 
If  
\be
|{\bf p}_{\pi 1} - {\bf p}_{\pi 2}| \lesssim \sigma_{p \pi}\,,   
\label{eq:pion}
\ee 
the energy and momentum conservation laws are still satisfied for different 
components of the wave function, while no entanglement occurs. Indeed, under 
this condition one can satisfy both equalities in eq.~(\ref{eq:mom}) 
simultaneously, which means that both neutrino mass eigenstates $\nu_1$ and 
$\nu_2$ can be produced with the muon having the same momentum ${\bf p}_\mu$, 
and therefore the correlation between the 4-momenta of the neutrino and the 
muon is absent. At the same time, by making use of relations (\ref{eq:mom}) 
we obtain from (\ref{eq:pion}) 
\be
|{\bf p}_{\nu 1} - {\bf p}_{\nu 2}| \lesssim \sigma_{p \pi}\,. 
\label{eq:coh2}
\ee 
This is the well known coherent production condition, which is a necessary 
condition for the observability of neutrino oscillations 
\cite{KayserQM,Rich:1993wu,Beuthe1,AS1}. 
It is essentially equivalent to the localization condition which requires 
that the pion decay region ({\it i.e.}\ the neutrino production region) be 
small compared to the oscillation length. Since the momenta of the two muon 
components may coincide, the integral in (\ref{eq:interf}) is unsuppressed. 
There is no kinematic entanglement, the coherence condition is 
satisfied at production, and if the neutrino detection is also coherent, 
the oscillations can be observed. The observability of neutrino oscillations 
does not depend on the muon detection.    

If $\sigma_{p \pi}\ll |{\bf p}_{\pi 1} - {\bf p}_{\pi 2}|$, the pion momentum 
is well defined and the neutrino and the muon are indeed produced in an 
entangled state. This reproduces approximately the extreme situation described 
in the beginning of this section (see eq.~(\ref{eq:exact}) and the following 
discussion)~\footnote{
For pion decay and known neutrino mass squared differences  
this is not realized even for the decay of a free pion. 
So, here we consider a {\it gedanken} situation with either very large 
$\Delta m^2$ or a very long lived and strongly delocalized 
nucleus instead of pion.}. 
In this case the observability of neutrino oscillations will depend on the 
conditions of detection of the muon and, of course, of the neutrino. 
If the muon is not detected, the disentanglement does not occur. The 
amplitudes corresponding to different neutrino mass eigenstates do not 
interfere and neutrino oscillations do not occur, independently of the 
conditions of neutrino detection. What actually happens is that the 
oscillations are averaged out due to the integration over the coordinate of 
the pion decay point in the production region. The coherence condition is not 
satisfied at neutrino production and cannot be restored at detection. 
This can also be seen from eq.~(\ref{eq:interf}). Since the pion momentum is 
well defined and energy-momentum conservation is exact, the muon components 
have different momenta: ${\bf p}_{\mu 1} - {\bf p}_{\mu 2} \neq 0$. Therefore 
the infinite-volume integration in (\ref{eq:interf}) leads to the 
vanishing result.  

The situation is different when  the muon is detected and therefore   
the muon momentum is measured with a finite resolution $\sigma_{p \mu}$. If 
\be 
|{\bf p}_{\mu 1} - {\bf p}_{\mu 2}| < \sigma_{p \mu}\,, 
\label{eq:coh}
\ee 
the two muon components $\mu(p_{\mu 1})$ and $\mu(p_{\mu 2})$ interfere 
constructively in the detector, and the produced neutrino state can be a 
coherent superposition of different neutrino mass eigenstates. 
Condition (\ref{eq:coh}) corresponds to the situation when the 
integration 
in (\ref{eq:interf}) is performed over the domain of the linear 
size $|\Delta {\bf x}_{\mu}| \lesssim 1/|\Delta {\bf p}_{\mu}|$, so that 
the integrand does not exhibit fast oscillations and the integral is is 
not suppressed. 
At the same time, condition (\ref{eq:coh}) implies disentanglement of the 
neutrino and the muon. Indeed, both $\nu_1$ and $\nu_2$ can be produced 
together with the same muon state which has the momentum uncertainty 
$\sigma_{p \mu}$.  

The observation of the muon in a finite-size detector provides its 
localization. This, together with the neutrino localization in its detection 
process, implies a localization of the decay region of the parent pion, 
{\it i.e.}\ of the neutrino production region. As a result, a washout of 
neutrino oscillations due to the averaging over the coordinate of the 
neutrino production point can be avoided. In other words, muon detection 
improves the localization of the pion decay region, thus improving the 
observability of neutrino oscillations.

If, however, $|{\bf p}_{\mu 1} - {\bf p}_{\mu 2}|\gg \sigma_{p \mu}$, the 
integral in (\ref{eq:interf}) is essentially zero, {\it i.e.} the 
interference is absent independently of the conditions of neutrino 
observation. In this case the pion and muon momenta have well-defined values, 
and therefore energy-momentum conservation fully determines the 4-momentum of 
the emitted neutrino. Because the neutrino propagates macroscopic distances 
and consequently is on the mass shell, its energy and momentum in turn 
uniquely determine its mass. Thus, in this case only a certain mass-eigenstate 
(rather than flavor-eigenstate) neutrino would be emitted, which would make 
neutrino oscillations impossible. 

The considerations presented in this section allow us to draw several  
general conclusions. 

1. The absence of entanglement (or the occurrence of disentanglement)  
corresponds to the situation when the coherence condition at neutrino 
production is satisfied, and {\it vice versa}, entanglement (and the absence 
of subsequent disentanglement) implies violation of the coherence condition 
at neutrino production.

2. For neutrino oscillations to be observable, two different 
localization 
conditions must be satisfied: apart from the localization of the neutrino 
detection process, either the parent particle or the recoil particle at 
neutrino production  
should be localized (the latter, together with the localization of the   
neutrino detection, will automatically localize the production process
as well). 

3. In the case of disentanglement which restores the observability 
of neutrino oscillations the event should consist of muon interaction in a 
relatively small space-time region (apart from the neutrino detection).
Just the fact that the muon interacts somewhere is not enough.

4. The kinematic disentanglement or the absence of entanglement occur when 
the energy and momentum uncertainties are introduced for the external 
particles which accompany neutrino production and detection. This is 
equivalent to describing the external particles by wave packets. Therefore, 
the treatment of neutrino oscillations in terms of possible entanglement and 
subsequent disentanglement can in a sense be considered as corresponding to 
the second QFT approach of sec.~\ref{sec:QFTem}, where the transition 
amplitudes are first calculated using plane waves (which implies exact 
energy-momentum conservation) and then are convoluted with momentum-space wave 
packets. If the coordinate-space wave packets are used from the very beginning, 
kinematic entanglement is completely irrelevant. The attempts to use the 
entanglement/disentanglement approach within the pure plane-wave framework 
are inconsistent.

5. The entanglement/disentanglement approach can describe certain, but not all, 
situations. It does not describe the most important case when the parent pion 
has a large enough momentum uncertainty, so that the entanglement does not 
arise at all. In this case neutrino oscillations can be observed even if 
the muon is not detected (disentangled). Neutrino oscillations can then be 
described in a consistent way without resorting to kinematic entanglement 
and in this sense the entanglement/disentanglement approach is irrelevant. 

6. The idea of making use of entanglement and subsequent disentanglement of 
the charged leptons and neutrinos stems from the desire to base the theory 
of neutrino oscillations on exact energy-momentum conservation, which 
is a rigorous law of nature. Let us stress once again, however, that the 
approximate conservation of mean energies and momenta in the wave 
packet approach does not contradict this exact conservation law. The  
approximate nature of conservation of the mean 4-momenta is related to the 
fact that the contributions to the 
overall energy and momentum balance coming from the surrounding particles, 
which localize the particles directly involved in the neutrino production and 
detection processes, are not taken into account. The directly involved 
particles form an open system. The energy and momentum conservation laws 
are, however, fulfilled exactly for contributions to the transition 
amplitudes coming from the individual momentum components of the wave 
packets of these particles \cite{Beuthe1} (sec.~\ref{sec:em}).

\subsection{\label{sec:recphase}The phases of the recoil particles}

In a number of papers \cite{Hamish,Hamish2,Kayser} it has been argued that  
kinematic entanglement implies that there are certain contributions of the 
phases $\phi_R$  of the recoil particles accompanying the neutrino production 
to the oscillation phase.  (In \cite{Boy1} it is assumed that the phase 
$\phi_R$ is small since the recoil is detected near the source.) 
Let us clarify this issue. Consider again the pion decay. 
For simplicity, we will consider only one spatial dimension, so that  
$$
p_\alpha = \{E_{\alpha}, k_{\alpha}\}, ~~~~
x_\alpha = \{t_\alpha, r_\alpha \} ~~~(\alpha = \pi, \mu, \nu)\,,
$$ 
keeping the possibility of the $\pm$ signs for momenta and coordinates 
(the generalization to the 3-dimensional case is straightforward). 
Suppose that the pion decays at the point $r=0$ at the time $t=0$, and that 
the muon and neutrino are detected at the points with 4-coordinates 
 $x_\mu = 
(t_\mu, r_\mu)$ and $x_\nu = (t_\nu, r_\nu)$, respectively. To the time of 
detection the state (\ref{eq:entang}) evolves into    
\be
|\mu(x_\mu) \nu (x_\nu)\rangle = 
\sum_i U_{\mu i}^* |\mu(p_{\mu i})\rangle |\nu_i(p_{\nu i})\rangle
e^{i (\phi_{\nu i}(x_\nu)+\phi_{\mu i}(x_\mu))}\,, 
\label{eq:entangev}
\ee
where 
\be
\phi_{\nu i}(x_\nu) = k_{\nu i} r_\nu - E_{\nu i} t_\nu\,, ~~~~
\phi_{\mu i }(x_\mu) = k_{\mu i} r_\mu - E_{\mu i } t_\mu\,.
\ee
The total phase difference between the two components of the state 
(\ref{eq:entangev}) corresponding to different neutrino mass 
eigenstates is 
\be
\Delta \phi = \Delta \phi_{\nu}(x_\nu) + \Delta \phi_{\mu}(x_\mu)\,, 
\label{eq:sumph}
\ee 
where 
\be
\Delta \phi_\nu  \equiv \phi_{\nu 2} - \phi_{\nu 1}\,, ~~~ 
\Delta \phi_\mu  \equiv \phi_{\mu 2} - \phi_{\mu 1}\,. 
\label{eq:Dphi}
\ee
According to (\ref{eq:sumph}) the probability of neutrino oscillations 
should depend on the \mbox{4-coordinate} of the point where the accompanying 
muon is detected. However, in all realistic setups the phase difference 
$\Delta \phi_{\mu}(x_\mu)$ between the components of the muon state  {\it is 
not measured}. Even if $\Delta \phi_{\mu}(x_\mu)$ were measured, this would 
not influence the neutrino oscillation pattern. Indeed, by itself the fact 
of the muon detection (irrespectively of whether or not the phase difference 
between its different components is measured) only establishes that {\it the 
neutrino produced in the source was the muon neutrino}. This
fixes the composition of this initial neutrino state in terms of the mass 
eigenstates. The subsequent change of the relative phase between the mass 
eigenstates (which leads to neutrino oscillations) is entirely determined by 
$\Delta m^2/2E_\nu$ and by the distance traveled by the neutrinos. 

Thus, we have a paradoxical situation: formally, the phase of the recoil 
particle enters into the expression for the entangled state. At the same 
time, the pattern of neutrino oscillations should not depend on this phase.
As we will see, the resolution of this paradox is trivial: 
{\it The recoil phase is negligible under the conditions of 
observability of neutrino  oscillations}. 
This phase is a part of the uncertainty of the oscillation phase due to 
the finite extension of the neutrino source. 

Let us discuss the muon phase difference in more detail. In general,   
\be
\Delta \phi_{\mu}(t_\mu, r_\mu) = \Delta k_\mu r_\mu - \Delta E_\mu t_\mu\,,  
\label{eq:phasemu}
\ee
where 
$\Delta k_\mu$ and  $\Delta E_\mu$ are the momentum and energy 
differences of the  muon components which correspond 
to two different neutrino mass eigenstates. From the mass shell relation 
$k^2_\mu = E^2_\mu - m_\mu^2$  we obtain  
\be
\Delta k_\mu = \Delta E_\mu \frac{E_\mu}{k_\mu}. 
\label{eq:disp}
\ee
Substituting this into~(\ref{eq:phasemu}) yields   
\be
\Delta \phi_{\mu}(x_\mu) =  
\Delta E_\mu \left(\frac{E_\mu}{k_\mu} r_\mu - t_\mu \right)  
= \frac{\Delta E_\mu}{v_\mu} \left(r_\mu -  v_\mu t_\mu \right), 
\label{eq:muphase}
\ee 
where $v_\mu = k_\mu/E_\mu$ is the muon velocity. If one assumes the muon to 
be a pointlike particle moving along a classical trajectory, the two terms in 
the brackets in eq.~(\ref{eq:muphase}) cancel each other, so that the 
contribution of the recoil to the phase difference~(\ref{eq:sumph}) 
vanishes: \mbox{$\Delta\phi_{\mu} = 0$.} 
Let us stress, however, that the fact that we assume all particles to have 
well-defined momenta implies that they are described by plane waves -- 
a notion which is just the opposite of pointlike particles. 
Thus, the argument presented above is actually controversial. Plane 
waves also bear other well known difficulties -- in particular, if the 
pion energy and momentum are well defined, the pion is completely delocalized
in space and time, so that the distances from the muon and neutrino 
production to detection points $x_\mu$ and $x_\nu$ cannot be defined.

In a consistent wave packet formalism, the equality $r_\mu = v_\mu t_\mu$ 
is satisfied (and the recoil phase vanishes) only at the center of the muon 
wave packet. In general, $r_\mu \ne v_\mu t_\mu$, but $|r_\mu - v_\mu t_\mu|$ 
cannot exceed significantly the spatial length of the muon wave packet  
$\sigma_{x \mu}$. Consequently, 
\be
|\Delta \phi_{\mu}| = \frac{\Delta E_\mu}{v_\mu} \left|r_\mu -  v_\mu 
t_\mu \right|
\lesssim  \frac{\Delta E_\mu}{v_\mu}\sigma_{x \mu}\,, 
\label{eq:dphase}
\ee
which is valid for all $r_\mu$ or $t_\mu$. The spatial length of the 
muon wave packet is related to its energy uncertainty as $\sigma_{x \mu} \sim  
1/ \sigma_{k \mu} = v_\mu/\sigma_{E \mu}$. Then, eq. (\ref{eq:dphase}) can be 
rewritten as 
\be
|\Delta \phi_{\mu}| \lesssim  \frac{\Delta E_\mu}{\sigma_{E\mu}}. 
\label{eq:cohhh}
\ee
This expression means that if the coherence condition  $\Delta E_\mu \ll 
\sigma_{E\mu}$, which is required for observability of neutrino oscillations, 
is fulfilled, the phase $\Delta\phi_{\mu}$ is negligible.  
(Recall that according to the coherence condition the energy and momentum 
uncertainties of the particles accompanying neutrino production and 
detection should be large enough, so that the momenta of different neutrino 
mass eigenstates can not be distinguished).

For neutrinos, instead of eq.~(\ref{eq:disp}) we have 
\be
\Delta k_\nu = \Delta E_\nu \frac{E_\nu}{k_\nu} - 
\frac{\Delta m^2}{2k_\nu}\,, 
\label{eq:dispnu}
\ee
and consequently the neutrino phase difference is 
\be
\Delta \phi_{\nu}(x_\mu) 
= \frac{\Delta E_\nu}{v_\nu} \left(r_\nu -  v_\nu t_\nu \right)
- \frac{\Delta m^2}{2k_\nu} r_\nu\,.  
\label{eq:nuphase}
\ee 
The key difference between $\Delta \phi_{\mu}$ and $\Delta \phi_{\nu}$  
originates from the fact that both components of the muon state have 
the same mass, whereas the components of the neutrino state 
have different masses  and therefore an additional contribution to 
$\Delta \phi_{\nu}$ appears due to the neutrino mass difference $\Delta m^2$
(cf. eqs.~(\ref{eq:muphase}) and (\ref{eq:nuphase})). 
The first term in (\ref{eq:nuphase}) can be estimated just as in the 
case of the muon wave packet: it vanishes at the center of the neutrino wave 
packet, is non-zero away from this point, but never exceeds significantly the 
spatial length of the neutrino wave packet 
$\sigma_{x \nu}\sim v_\nu/\sigma_{E\nu}$, where $\sigma_{E\nu}$ is the 
energy uncertainty characterizing the neutrino state. Therefore
\be
\frac{\Delta E_\nu}{v_\nu} \left|r_\nu -  v_\nu t_\nu \right|
\lesssim
\frac{\Delta E_\nu}{\sigma_{E \nu}}\,.   
\label{eq:dnuphase}
\ee

Combining eqs.~(\ref{eq:sumph}), (\ref{eq:muphase}) and (\ref{eq:nuphase}) 
we obtain for the total phase difference 
\be
\Delta \phi = \frac{\Delta E_\mu}{v_\mu} \left(r_\mu -  v_\mu t_\mu \right)
+ \frac{\Delta E_\nu}{v_\nu} \left(r_\nu -  v_\nu t_\nu \right) 
- \frac{\Delta m^2}{2k_\nu} r_\nu\,.  
\label{eq:dpsigma}
\ee
{}From eqs.~(\ref{eq:dphase}) and~(\ref{eq:dnuphase}) it follows that the 
first two terms in~(\ref{eq:dpsigma}) are negligible when the energy 
differences of the different components of the muon and neutrino states 
are small compared to their respective energy uncertainties, {\it i.e.}\
when the coherent neutrino production condition is satisfied, as advertised.  

In a complete QFT approach the phase of parent particle, 
that is, the pion phase $\Delta \phi_{\pi}(x_\pi)$ in our example, also 
should be taken into account. Now  instead of (\ref{eq:sumph}), we have %
\footnote{Equation~(\ref{eq:sump}) can be obtained by 
considering the amplitudes in eq. (\ref{eq:A2}). } 
\be
\Delta \phi = - \Delta\phi_{\pi}(x_\pi) + \Delta\phi_{\nu}(x_\nu) + 
\Delta\phi_{\mu}(x_\mu), 
\label{eq:sump}
\ee 
where the minus sign in front of $\Delta\phi_{\pi}$ is related to the fact 
that the pion is destroyed, whereas the muon and the neutrino are produced. 
Here $x_\pi=(t_\pi, r_\pi)$, and all distances and times are counted from 
the point of the pion decay.  
The expression for $\Delta \phi_{\pi}$ is similar to that for the muon, and 
therefore for the total phase difference we obtain 
\be
\Delta \phi=
-\frac{1}{v_\pi}({r}_{\pi } - {v}_{\pi} t_{\pi }) \Delta E_{\pi}+
\frac{1}{v_\mu}({r}_{\mu }- {v}_\mu t_{\mu }) \Delta E_\mu + 
\frac{1}{v_\nu}({r}_{\nu } - {v}_\nu t_{\nu })\Delta E_\nu - 
\frac{\Delta m^2}{2k_\nu} {r}_{\nu }\,.
\label{eq:phase2}
\ee
Equation~(\ref{eq:phase2}) is equivalent to eq.~(39) of Dolgov {\it et al.} 
\cite{Dolgov}, where the dependence of the oscillation phase on the  
4-coordinate of the muon was interpreted as an indication of the EPR-type 
correlation in neutrino oscillations in the case when both muon and neutrino 
are detected. While we agree with that equation, we disagree with its 
interpretation given in \cite{Dolgov}. 

Indeed, the pion phase difference can be estimated similarly to that for the 
muon. As a result,  we find that each of the first three terms on the right 
hand side of eq.~(\ref{eq:phase2}) is smaller than or of the order of 
$\Delta E_\alpha/\sigma_{E \alpha}$ ($\alpha = \pi,\,\mu,\,\nu)$.  
The condition of coherent neutrino production requires that all the 
involved energy differences be small compared to the corresponding energy 
uncertainties (otherwise one would have been able to determine which neutrino 
mass eigenstate was emitted). Hence, if the coherent production condition is 
satisfied, the first three terms on the right hand side of 
eq.~(\ref{eq:phase2}) are small compared to unity and can 
be neglected; the remaining (last) term just gives the standard oscillation 
phase. We therefore conclude that the condition of coherent production  
eliminates the dependence of the oscillation phase 
on the 4-coordinate of the muon, and no EPR-like correlations arise. 
On the other hand, if the coherence 
condition is violated, neutrino oscillations are unobservable.

The effect of the recoil particle can be understood from fig.~2.
If the muon is not detected or the duration of the detection process exceeds 
$\tau_\pi$, the oscillation diagram coincides with the one shown in the 
figure. In this case the oscillation pattern is not affected by the recoil 
particle: it is determined by the localization of the pion and by the 
neutrino parameters. If the  muon detection proceeds during a shorter time 
interval, $t_{det} <  \tau_\pi$, the muon (green) band will be
narrower. Correspondingly, the production (overlap) region will be
shorter. That, in turn, will reduce  the width of the neutrino wave packet 
and consequently will improve the production coherence.

\subsection{\label{sec:EPR} Recoil particle phase 
and the EPR paradox}

\begin{figure}
  \begin{center}
\includegraphics[height=5.2cm,width=12.0cm]{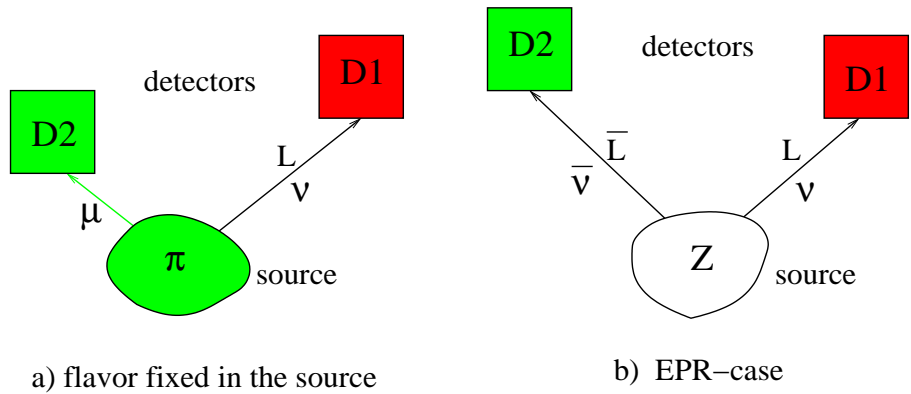}
  \end{center}
\caption{Two-detector neutrino oscillation experiments: Neutrino 
production along with $\mu$ in a pion decay (a) and 
$\nu\bar{\nu}$ production in a $Z^0$ decay (b). EPR correlation is 
present only in the second case.}
  \label{fig:2det}
\end{figure}

Let us compare the situation described in the previous subsection 
with the case of neutrinos produced in $Z^0$ decays. Here the 
conditions for the EPR paradox are realized and the phases of both 
products of the decay are important for the oscillations \cite{SZ}. 
The neutrino state produced in the $Z^0$ decay is given by 
\be
|\nu_Z \rangle  = \frac{1}{\sqrt{3}} \sum_i | \bar{\nu}_i \nu_i \rangle\,.  
\label{eq:nuz}
\ee
The neutrino mass is not determined; all mass eigenstates are produced 
with the same probability amplitude. The  mass eigenstates are entangled. 
Let us consider a {\it gedanken} experiment 
with two neutrino detectors sensitive to the neutrino mass. 
If the first  detector observes a mass-eigenstate component $\nu_i$, then the 
second detector should see the  antineutrino $\bar{\nu}_i$ of the same mass,  
independently of the  distance between the detectors and source. 

The situation with the flavor neutrino states is more complicated,   
since these states are not eigenstates of the free Hamiltonian, 
{\it i.e.} they may oscillate. The $Z^0$-boson interactions are flavor blind 
and therefore all three flavors 
are produced with the same probability amplitude. Indeed, from $|\nu_\alpha
\rangle =\sum_i U_{\alpha i}^*|\nu_i\rangle$ it follows that~(\ref{eq:nuz}) 
can be rewritten as  
\be
|\nu_Z \rangle  = \frac{1}{\sqrt{3}} 
\sum_{\alpha = e, \mu, \tau} | \bar{\nu}_\alpha \nu_\alpha \rangle 
\label{eq:nuzf}
\ee  
and is independent of the elements of the leptonic mixing matrix 
$U_{\alpha i}$. The flavor states correlate at the production: if both 
detectors are located near the production region and, e.g., $\nu_e$ is 
detected  by the first  detector, then in the second detector $\bar{\nu}_e$ 
should be observed. If the detectors is situated at large 
enough distance from the source, so that neutrinos have time to oscillate, 
the flavor correlations will depend on the distance between the source and the 
detectors (see fig.~\ref{fig:2det}). If one of the detectors observes, e.g., 
$\nu_{\mu}$, 
this will fix the flavor composition of the antineutrino state observed by 
the second  detector. It will be determined by the sum of the phase 
differences acquired by neutrinos and antineutrinos on their ways from the 
source to the detectors \cite{SZ}:  
\be
\Delta \phi = \Delta \phi_\nu + \Delta \bar{\phi}_\nu.
\ee 
The oscillation picture corresponds to the situation as if one of the 
detectors was a source of the neutrino and the other one a detector. 

The key condition for the observability of neutrino oscillations is 
that the flavor composition of the neutrino state be measured in two different 
space-time points (fig.~\ref{fig:2det}). If the neutrino flavor is fixed in one 
space-time point, then the oscillatory pattern can be observed in another 
space-time point. 
In the case under consideration, the source ($Z^0$ decay) does not fix 
the neutrino flavor, and therefore two flavor detectors are necessary.  
The  difference of this EPR setup from neutrino production in, e.g.,    
$\pi\to \mu\nu$ decay is that in the latter case the flavor of the neutrino 
state is fixed at production independently of the place where the muon 
is detected, and even of whether the muon is actually detected or not.

\section{\label{sec:concl}Conclusions}

In this paper we have considered the related issues of entanglement, 
energy-momentum conservation and possibility of consistent description 
of neutrino oscillations in the $S$-matrix formalism.   

1. We have shown that neutrino oscillations, being intrinsically a finite 
space-time phenomenon, can be consistently described in the $S$-matrix 
formalism provided that the correct physics conditions of the oscillation 
setup are imposed. The key condition is the existence of two finite space-time 
interaction regions: the production region and the detection region. The 
$S$-matrix formalism should be adjusted correspondingly. In the context of 
QFT this adjustment requires using wave packets for the external particles, 
i.e. the particles accompanying neutrino production and detection. This 
allows one to describe their localization, and consequently, define the 
neutrino production and detection regions. The wave packets encode 
information about the interactions of the particles 
accompanying neutrino production and detection 
with the external (to the neutrino processes) system which provides 
the localization of these particles.

Instead of using wave packets for the external particles, one can employ 
plane waves amended by finite-interval space-time integrations over the 
4-coordinates of the neutrino production and detection points. 
This can be considered as an approximation to the description of neutrino 
oscillations in a realistic wave packet framework. 
Since finite pieces of plane waves are in fact wave packets, 
this approach can actually be considered as a primitive wave packet one. 

2. We have demonstrated the equivalence of two approaches to calculating  
the oscillation amplitude: (i) Using configuration-space wave packets  
for the external particles involved in neutrino production and detection. 
(ii) Using  plane waves for the external particles and the subsequent 
convolution of the obtained amplitudes with the wave packets of the external 
particles in the momentum representation. These two approaches allow different 
physics interpretation. The first one underlines the approximate nature of 
conservation of the mean energies and momenta of the involved particles. In 
the second approach one can speak of exact energy-momentum conservation for 
the individual momentum components of the wave packets. However, these  
individual components appear only at an intermediate stage of calculations, 
and at this stage the oscillations cannot be obtained. The 
required localization of the neutrino production and detection processes and, 
consequently, the oscillations are only obtained after the convolution of the 
plane-wave amplitudes with the momentum-space wave packets of the external 
particles which encode the information about their localization.

3. The assumption that particles involved in neutrino production processes 
have well-defined energies and momenta which satisfy exact conservation laws 
leads to kinematic entanglement of the neutrinos and the accompanying 
particles (e.g., charged daughter particles in decays). This entanglement 
destroys coherence of a neutrino state and therefore oscillations become 
unobservable unless energy and momentum uncertainties are introduced for the 
recoil and/or parent particles. To observe the oscillations, one should 
arrange localization in two different sites of the experimental setup: at 
neutrino production (the parent and/or recoil particles should be localized) 
and at detection (the neutrino absorption process should be localized). 
This provides the requisite energy and momentum uncertainties which make 
neutrino oscillations possible.   

The conditions for entanglement and disentanglement are related to the 
coherence condition at neutrino production and detection. In the case when the 
production process is delocalized, energy-momentum conservation leads 
to kinematic entanglement of neutrinos and recoil particles but prevents 
coherent emission of different neutrino mass eigenstates and therefore 
makes the oscillations unobservable. Disentanglement of neutrinos and recoil 
particles through the detection of the latter can restore the coherence of the 
emitted neutrino state. If, however, the localization of the parent particle is 
good enough, so that the coherence condition at the neutrino production 
is satisfied, no entanglement occurs, and the corresponding description 
is irrelevant. A complete and consistent description of the 
oscillations is then possible without resorting  to the 
entanglement/disentanglement approach.

4. We show that, for processes induced by charged-current weak interactions, 
the phases of the recoil particles (e.g., of charged leptons associated with 
neutrino production or detection) are irrelevant for neutrino oscillations. If 
the coherent production condition is fulfilled, the contribution of the recoil 
particle's phase to the oscillation phase is negligible and, in fact, is 
included in the uncertainty related to the finite extension of the source. No 
EPR-like effects arise. This differs from the case of neutrino production by 
neutral currents (e.g. in $Z^0$ decay). Here the flavor of the neutrino state 
is not fixed at the production point, and the oscillations can only be 
observed in a two detectors experiment, when one of the detectors fixes the 
flavor composition of the neutrino state and the other one observes the 
oscillation pattern in the antineutrino state (or {\it vice versa}). In 
contrast to this, in charged current processes the neutrino flavor is 
already fixed at production, and therefore neutrino oscillations do not 
depend on the evolution of the recoil particles.

\appendix
\renewcommand{\theequation}{\thesection\arabic{equation}}
\appsection
\renewcommand{\thesection}{\Alph{section}}
\section*{Appendix \Alph{section}:
Finite observation time approach of~\cite{Boy1,Boy2}}

In all the situations considered in~\cite{Boy1} the physics setups did not 
correspond to those of realistic neutrino oscillation experiments. 
The authors used the plane wave formalism but did not introduce the 
correct limits of integration over the 4-coordinates of the neutrino 
emission and absorption points, which is mandatory in that case. 
Three different computations of the oscillation probabilities were presented 
in~\cite{Boy1}:

1. In the case called ``the $S$-matrix result'', the authors consider 
a single region of neutrino interactions and perform integrations over the 
4-coordinates of the production and detection vertices in the same infinite 
intervals. This corresponds to a scattering rather than oscillation setup 
(fig.~\ref{fig:scheme1}). Such an integration leads to averaging over all 
possible baselines $L$ from 0 to $\infty$, and consequently, the oscillations 
are averaged out. 

2. In the second case the authors again took a single interaction region and 
used infinite spatial integration, but performed time integration over a 
finite interval from $t^i$ to $t^f$. The time ordering has been imposed, so 
that the charged lepton associated with neutrino production is observed near 
the production point earlier than the charged lepton associated with neutrino 
detection which is observed near the neutrino absorption point. The 
integrations over the production and detection times $t_1$ and $t_2$ 
are performed over the same interval $(t^i, t^f)$ with $t_1 \leq t_2$.  
This would correspond to the averaging (integration) over the baselines $L$ 
from $0$ to $v (t^f - t^i)$, where $v$ is the neutrino velocity. Now, due to 
the finite time integration, the oscillatory terms appear in the expression 
for the transition probability, {\it i.e.}\ the oscillations are not averaged 
out. Thus, the results obtained in this case differ substantially from those 
in the previous case. From this the authors conclude that the $S-$matrix 
formalism is ill-suited to describe neutrino oscillations. However, both 
results are incorrect in the sense that they do not correspond to a realistic 
oscillations setup. As is discussed in sec.~\ref{sec:pw} of the present paper, 
with the correct setup the $S$-matrix formalism is perfectly adequate for 
describing neutrino oscillations.   

3. In the third computation the authors assumed disentanglement of the 
neutrino and charged lepton in the final state of the production process and 
introduced two different intervals of time integrations. The integration over 
$t_1$ is performed from $t = 0$ (the moment of neutrino production) to 
$t_{S}$ --  the time of detection (disentanglement) of the accompanying 
charged lepton. This detection essentially occurs in the neutrino source, so 
that $v t_S \ll l_\nu$, where $l_\nu$ is the neutrino oscillation length. The 
time $t_{S}$ is of the order of or larger than the production interval 
$t^S_f - t^S_i $ considered in our paper (see sec.~\ref{sec:pw}). The second 
integration is from $t_{S}^f$ to $t_D^f$ (in our notation). This interval is 
simply attached to the production interval, which is definitely incorrect for 
the long-baseline oscillation setup, provided that one uses plane waves. 
This integration corresponds to the situation when the wave functions of 
the external particles participating in neutrino detection are not localized 
inside the detector but are distributed uniformly all the way from the 
neutrino production region ($t\sim t_{S}^f$) to the detection region 
($t\sim t_{D}^f$). That is, here one deals with a detector that is as long as 
the baseline itself. It is this time integration over the interval $t_{S}^f - 
t_D^f$ that leads in~\cite{Boy1} to the oscillation probability which is 
actually the integral of the standard neutrino oscillation probability.\\

\appsection
\renewcommand{\thesection}{\Alph{section}}
\section*{Appendix \Alph{section}:
Entanglement and the oscillation length}

In refs, \cite{Hamish,Kayser} two-body decays $P \rightarrow \nu + R$, 
were considered, where $P$ is the parent particle and $R$ is the ``recoil''.   
The total phase difference between the $\nu_1$- and $\nu_2$-containing  
components of the entangled state $|\nu R\rangle$ is 
\be
\Delta \phi = \Delta \phi_{\nu} + \Delta \phi_{R}. 
\ee 
In turn, the phase differences $\Delta \phi_{\nu}$ 
and $\Delta \phi_{R}$ each 
have two contributions:  the contribution from the difference of energies 
and the contribution from the difference of 
momenta, so that 
\be
\Delta \phi  = (\Delta k_R r_R - \Delta E_R t_R)  + 
(\Delta k_\nu r_\nu - \Delta E_\nu t_\nu)\,,  
\label{eq:eqtot1}
\ee 
where 
\bea
\Delta E_R & \equiv & E_{R2} - E_{R1},~~~\Delta E_\nu \equiv E_{\nu 2} 
- E_{\nu 1}\,,  
\nonumber\\
\Delta k_R & \equiv & k_{R 2} - k_{R 1},~~~
\Delta k_\nu \equiv k_{\nu 2} - k_{\nu 1}\,. 
\eea
In sec.~\ref{sec:recphase} we have shown that 
\be
|\Delta k_R r_R - \Delta E_R t_R| \;\lesssim\; 
\frac{\Delta E_R}{\sigma_{E_R}}\,,
\label{eq:eqzero}
\ee
where $\sigma_{E_R}$ is the energy uncertainty of the state of the recoil 
particle. As shown in sec.~\ref{sec:recphase}, the coherent neutrino 
production condition, which is a necessary condition for the observability of 
neutrino oscillations, requires, in particular, $\Delta E_R/\sigma_{E_R}\ll 1$. 
Therefore eq.~(\ref{eq:eqzero}) means that under the coherence condition 
one has $\Delta \phi_{R} \ll 1$, {\it i.e.}\ the term in the first brackets 
in~(\ref{eq:eqtot1}) can be neglected. Hence the total phase 
difference is simply reduced to the phase difference of the two neutrino mass 
eigenstates, $\Delta \phi \approx \Delta \phi_{\nu}$ which, in turn, leads to 
the standard oscillation phase and length.  This result does not depend on the 
distance and time of detection of the recoil particle and it does not depend 
on whether or not the energy-momentum conservation is applied. Note that in 
ref.~\cite{Kayser} the conclusion that the oscillation phase takes its 
standard value was obtained assuming that the neutrino and the recoil 
particles move along classical trajectories, i.e. are in fact pointlike. This 
is not a consistent assumption because the consideration 
in~\cite{Hamish,Kayser} was based on the plane wave approach, and the notion 
of pointlike particles is just the opposite of that of plane waves.

Most of the complications encountered in refs.~\cite{Hamish,Kayser} originate 
essentially from the following grouping of the terms in the expression for the 
total phase difference (\ref{eq:eqtot1}): 
\be
\Delta \phi  = 
(\Delta k_R r_R + \Delta k_\nu r_\nu) 
- (\Delta E_R t_R + \Delta E_\nu t_\nu)\,,   
\label{eq:phasetot}
\ee
which differs from that in (\ref{eq:eqtot1}). Following~\cite{Hamish}, 
we shall  consider the problem in the rest frame of the parent particle $P$ 
(the generalization to the case of $P$ decay in flight is trivial). 
In~\cite{Hamish} exact energy-momentum conservation was applied and it was 
assumed that 
\be
t_R =  t_\nu \equiv t \,, 
\label{eq:assump}
\ee 
{\it i.e.}\, that the moments of time at which the recoil and the neutrino are 
observed coincide in the rest frame of $P$. In principle, this is possible, 
in general this is not justified, and in practice this is never realized.  
Under condition (\ref{eq:assump}), for the expression in the second 
brackets of eq.~(\ref{eq:phasetot}) one obtains   
\be 
\Delta E_R t_R + \Delta E_\nu t_\nu=(\Delta E_R + \Delta E_\nu)t = 0\,. 
\ee
The second equality here holds because, due to energy conservation. one has 
$E_{R2} + E_{\nu 2} = E_{R 1} + E_{\nu 1} = m_P$, where $m_P$ is the mass of 
the parent particle, so that $\Delta E_R=-\Delta E_\nu$. Consequently, 
\be
\Delta \phi (t) = \Delta k_R r_R + \Delta k_\nu r_\nu = 
\Delta k_\nu (r_\nu - r_R)\,,  
\label{eq:phase-p}
\ee
where it has been taken into account that due to the momentum conservation 
$k_{R i}=-k_{\nu i}$ $(i = 1, 2)$ and therefore, $\Delta k_R = - \Delta k_\nu$. 
Then in \cite{Hamish} it was concluded that according to eq.~(\ref{eq:phase-p})
one should observe the oscillation pattern with the oscillation length 
\be
l_{R\nu} = \frac{2\pi}{\Delta k_\nu}\,. 
\ee
Here $\Delta k_\nu$ is determined by the kinematics of the decay and depends 
on the masses of the particles involved. Consequently, $l_{R\nu}$ differs from 
the standard oscillation length and turns out to be process-dependent. 

Subsequently, in ref.~\cite{Hamish2} the length  $l_{R\nu}$ was called the 
``separation wavelength'' which  determines the oscillatory pattern with 
respect to the separation $D \equiv r_\nu - r_R$ between the neutrino and the 
recoil. Yet another oscillation length has been introduced -- ``the oscillation 
length relative to the decay point''. The phase (\ref{eq:phase-p}) can be 
rewritten in terms of the distance traveled by the neutrino in the standard 
form 
\be
\Delta \phi (t) = \Delta k_\nu \left(1 - \frac{r_R}{r_\nu} \right) r_\nu = 
2\pi  \frac{r_\nu}{l_\nu},
\label{eq:phase-nu}
\ee
where 
\be
l_\nu \equiv  l_{R\nu}\frac{1}{1 - \frac{r_R}{r_\nu}}. 
\ee
Since $r_R$ and $r_\nu$ are measured in the same moment of time $t$ in the 
rest frame of the parent particle, for pointlike neutrinos and recoil 
particles these two distances are related, and their ratio is  determined 
by the velocities of the involved particles. It is then straightforward to 
show that $l_\nu$ coincides with the standard oscillation length. 

Our point is that all these complications and the introduction of the 
``separation'' oscillation length have no physical meaning. If the 
coherent production condition is satisfied, the additional ``recoil'' term 
$\Delta k_\nu r_R$ in the expression for $\Delta \phi (t)$ 
(see (\ref{eq:phase-p})) is essentially nothing but the contribution to 
the oscillation phase coming from the difference of neutrino energies:  
$\Delta k_\nu r_R \approx \Delta E_\nu t $. 
Indeed, we have 
\be
\Delta k_\nu r_R = - \Delta k_R r_R \approx - \Delta E_R t =  \Delta 
E_\nu t\,,  
\ee
where the first equality follows from the momentum conservation, the second 
(approximate) equality -- from eq.~(\ref{eq:eqzero}) and the 
coherence condition $\Delta E_R/\sigma_{E_R}\ll 1$ (which imply $\Delta k_R 
r_R \approx \Delta E_R t$), and the last one  is due to energy conservation. 
As a result, the expression for the oscillation phase takes the usual 
form 
\be
\Delta \phi  = \Delta k_\nu r_\nu - \Delta E_\nu t_\nu   
\label{eq:phasetot2}
\ee
with only the neutrino contribution present. 

\vspace*{2mm}
{\em Acknowledgments}. The authors are grateful to T. Goldman, B. Kayser,  
J. Kopp and H. Robertson for useful discussions and to D. Boyanovsky for 
a correspondence on ref.~\cite{Boy1}. This work was supported in part by the 
Sonderforschungsbereich TR 27 of the Deutsche Forschungsgemeinschaft (EA), and  
by the Alexander von  Humboldt foundation (AYS).

\end{document}